\documentclass[12pt]{article}
\usepackage{amssymb}
\usepackage{amsmath}
\usepackage{epsfig}
\usepackage{psfrag}
\usepackage{latexsym}
\usepackage{euscript}
\usepackage{bbold}

\usepackage{rotating}
\usepackage{verbatim}
\usepackage{pstricks}
\usepackage{array}
\def\gs{\mathrel{
   \rlap{\raise 0.511ex \hbox{$>$}}{\lower 0.511ex \hbox{$\sim$}}}}
\def\ls{\mathrel{
   \rlap{\raise 0.511ex \hbox{$<$}}{\lower 0.511ex \hbox{$\sim$}}}}

\newcommand{\ba}{\begin{array}{c}}
\newcommand{\baz}{\begin{array}{cc}}
\newcommand{\bad}{\begin{array}{ccc}}
\newcommand{\ea}{\end{array}}

\newcommand{\be}{\beta}

\def\beq{\begin{equation}}
\def\eeq{\end{equation}}
\def\bea{\begin{eqnarray}}
\def\eea{\end{eqnarray}}
\def\bet{\begin{tabular}}
\def\eet{\end{tabular}}
\def\bes{\begin{subequations}\bea}
\def\ees{\eea\end{subequations}}

\def\be{\begin{equation}}
\def\ee{\end{equation}}

\def\bea{\begin{eqnarray}}
\def\eea{\end{eqnarray}}
\def\nn{\nonumber}

\def\be{\begin{equation}}
\def\ee{\end{equation}}
\def\bc{\begin{center}}
\def\ec{\end{center}}
\def\bea{\begin{eqnarray}}
\def\eea{\end{eqnarray}}

\def\nn{\nonumber}

\catcode`@=11
\def\marginnote#1{}
\newcount\hour
\newcount\minute
\newtoks\amorpm
\hour=\time\divide\hour by60
\minute=\time{\multiply\hour by60 \global\advance\minute by-\hour}
\edef\standardtime{{\ifnum\hour<12 \global\amorpm={am}%
        \else\global\amorpm={pm}\advance\hour by-12 \fi
        \ifnum\hour=0 \hour=12 \fi
        \number\hour:\ifnum\minute<10 0\fi\number\minute\the\amorpm}}
\edef\militarytime{\number\hour:\ifnum\minute<10 0\fi\number\minute}
\def\draftlabel#1{{\@bsphack\if@filesw {\let\thepage\relax
   \xdef\@gtempa{\write\@auxout{\string
      \newlabel{#1}{{\@currentlabel}{\thepage}}}}}\@gtempa
   \if@nobreak \ifvmode\nobreak\fi\fi\fi\@esphack}
        \gdef\@eqnlabel{#1}}
\def\@eqnlabel{}
\def\@vacuum{}
\def\draftmarginnote#1{\marginpar{\raggedright\scriptsize\tt#1}}
\def\draft{\oddsidemargin 0.0truein
        \def\@oddfoot{\sl preliminary draft \hfil
        \rm\thepage\hfil\sl\today\quad\militarytime}
        \let\@evenfoot\@oddfoot \overfullrule 3pt
        \let\label=\draftlabel
        \let\marginnote=\draftmarginnote
   \def\@eqnnum{(\theequation)\rlap{\kern\marginparsep\tt\@eqnlabel}%
\global\let\@eqnlabel\@vacuum}  }
\catcode`@=12

\begin{document}
\begin{titlepage}
\vspace*{-1cm}
\phantom{hep-ph/***}
\hfill{DFPD-2011/TH/11}\\
\vskip 2.5cm
\begin{center}
\mathversion{bold}
{\Large\bf Discrete Flavour Symmetries in Light of T2K}
\mathversion{normal}
\end{center}
\vskip 0.5  cm
\begin{center}
{\large Reinier de Adelhart Toorop}~\footnote{e-mail address: reintoorop@nikhef.nl}
\\
\vskip .2cm
Nikhef Theory Group
\\
Science Park 105, 1098 XG Amsterdam, The Netherlands
\\
\vskip .2cm
and
\\
{\large Ferruccio Feruglio}~\footnote{e-mail address: ferruccio.feruglio@pd.infn.it},
{\large Claudia Hagedorn}~\footnote{e-mail address: claudia.hagedorn@pd.infn.it}
\\
\vskip .2cm
Dipartimento di Fisica `G.~Galilei', Universit\`a di Padova
\\
INFN, Sezione di Padova, Via Marzolo~8, I-35131 Padua, Italy
\end{center}
\vskip 0.7cm
\begin{abstract}
\noindent
We show that a non-vanishing angle $\theta_{13}$ of order 0.1 can be predicted in the framework of discrete
flavour symmetries. We assume that left-handed leptons transform as triplets under a group
 $G_f$ which is broken in such a way that neutrino and charged lepton sectors remain invariant under the subgroups
$G_\nu$ and $G_e$ of $G_f$, respectively. 
In this limit mixing angles and the Dirac $CP$ violating
phase $\delta_{CP}$ are determined. By choosing $G_f=\Delta(6 n^2)$ ($n=4,8)$, $G_\nu=Z_2\times Z_2$
and $G_e=Z_3$ we find $\sin^2\theta_{13}=0.045(0.011)$ for $n=4(8)$. At the same time $\theta_{23}$ and $\theta_{12}$  remain close to their experimental best fit values, particularly in the case $n=8$,
where $\sin^2\theta_{23}\approx 0.424$ and $\sin^2\theta_{12}\approx 0.337$. $\delta_{CP}$ is predicted
to be $0$ or $\pi$ so that $CP$ is conserved in our examples. 
\end{abstract}
\end{titlepage}
\setcounter{footnote}{0}
\vskip2truecm

\section{Introduction}

Neutrino oscillation experiments, interpreted in the framework of three active neutrino species,
have shown that the mixing angle $\theta_{13}$ is considerably smaller than the other two. Until recently
$\theta_{13}$ was actually compatible with being zero at the $2 \sigma$ level. 
 In June 2011 the T2K collaboration reported indication of electron neutrino appearance from a muon neutrino beam of energy about 0.6 GeV produced at J-PARC,
295 km away from the detector \cite{abe11}. This excludes the hypothesis $\theta_{13}=0$ at the level of
 2.5$\sigma$ and favors $\theta_{13}$ around $0.17\div 0.19$
\footnote{Angles are given in radian.}, not far from the upper limits set by CHOOZ \cite{chooz} and by MINOS
\cite{minos}. By itself such an indication is not conclusive but it adds in an interesting way to other previous hints suggesting
a non-vanishing reactor mixing angle in that range. In particular a tension between the 
values of the oscillation parameters extracted from KamLAND and from solar neutrinos
is alleviated for $\theta_{13}\approx 0.1$ \cite{theta13nonzero08}. A recent global analysis of the
 data \cite{fogli} provides evidence for nonzero $\theta_{13}$ at the 3$\sigma$ level
\be
\sin^2\theta_{13}=0.021(0.025)\pm0.007~~~(1\sigma)~~~.
\label{fogli}
\ee
The central value 0.15 (0.16) of $\theta_{13}$ depends on the assumed reactor antineutrino flux, with the results from the new flux estimate \cite{flux} shown in parenthesis. The two other angles are (at $1\sigma$ level)
\be
\sin^2\theta_{23}=0.42^{+0.08}_{-0.03}~~~,~~~~~~~~~~~~\sin^2\theta_{12}=0.306(0.312)^{+0.018(0.017)}_{-0.015(0.016)}~~~.
\ee
Notice that the difference between new and old fluxes is only relevant for $\theta_{12}$ and $\theta_{13}$.

Finding a consistent and economic explanation of fermion masses and mixing
angles is one of the main open problems in particle physics today. Given the key role traditionally
played by symmetries in understanding particle properties, flavour symmetries have captured considerable attention in this context. The peculiar mixing in the lepton sector,
with two large angles and a small one, significantly close to simple patterns
such as $\sin^2\theta_{23}=1/2$, $\sin^2\theta_{12}=1/3$ and $\sin^2\theta_{13}=0$, 
has revived the interest in discrete groups, that naturally incorporate this kind
of patterns. The mentioned pattern is called tribimaximal (TB) mixing \cite{HPS} and 
has received much attention in the last years. Many efforts have been made to reproduce it in concrete models and one of the simplest ways is via a non-trivially broken flavour symmetry based on a small discrete group, such as $A_4$ and  $S_4$, respectively, see \cite{reviews} for reviews. TB mixing is in general obtained
in a certain limit of the theory in which corrections
are neglected. Deviations thereof are expected
to arise from several sources and  non-vanishing $\theta_{13}$ is predicted.
However, these corrections tend to affect all mixing angles by a similar amount, and, given the
good agreement between the predicted and the observed value of the solar mixing angle,
only small corrections, up to $0.03$, are admissible. As a consequence, $\sin \theta_{13}$ is also
expected to be of order $0.03$ and thus not compatible with the result in eq. (\ref{fogli}).
The same arguments apply to other mixing patterns that
have been derived from a non-trivial breaking of other discrete groups, such as patterns with $\theta_{12}$
given in terms of the golden ratio \cite{gr} or given as $\sin^2 \theta_{12}=1/4$ \cite{hexa}.
Also in these cases the good agreement between predicted and observed value of the solar angle
suggests that corrections should be small, of the order of few percent.

Several attempts have been made to explain largish $\theta_{13}$ in a framework predicting TB mixing at
leading order (LO): the introduction of corrections leaving one row or column of the mixing matrix unchanged \cite{TBmod,zee};
the addition of scalars transforming as non-trivial singlets in $A_4$ models \cite{ma} leading to an explicit
breaking of the $\mu\tau$ exchange symmetry of the neutrino sector (in the charged lepton mass basis)
\footnote{For a similar model with the flavour symmetry $S_4$, see \cite{mpp_2011}.} or
the assumption that the different symmetry breaking parameters associated with the neutrino
and charged lepton sectors, respectively, are significantly different in size \cite{lin}. In other models \cite{meloni} bimaximal mixing is derived as mixing pattern at LO
which requires sizable corrections to the solar mixing angle and easily leads to large $\theta_{13}$ as well.
However, in this case the atmospheric mixing angle has to be protected from too large corrections and
it becomes difficult to precisely predict a particular value for the mixing angles.
 
 In the present note we change perspective and show that it is indeed possible to derive
mixing patterns with $\theta_{13} \neq 0$ at LO from discrete flavour symmetries $G_f$. We break $G_f$
in a non-trivial way so that the neutrino and the charged lepton sectors are (separately) invariant under
 two different subgroups of the original group $G_f$.
We show two examples, based on the flavour groups $\Delta(6n^2)$, with $n=4$ and $n=8$ respectively,
 in which not only $\theta_{13}$ of the correct size is predicted, but also $\theta_{23}$ and $\theta_{12}$ 
 are close to their experimental best fit values, e.g. for $n=8$ we get $\sin^2\theta_{23}\approx 0.424$ and $\sin^2\theta_{12}\approx 0.337$.

 \section{Framework}

In our approach the theory is invariant under
a discrete flavour group $G_f$ under which the three generations of SU(2)$_L$ lepton doublets $l$ transform
as a faithful three-dimensional irreducible representation. The group $G_f$ can be a symmetry of the full Lagrangian 
or just an accidental one arising in some LO approximation, for instance by neglecting operators
of high dimensionality. The lepton mixing matrix $U_{PMNS}$ is determined by the residual symmetries of the neutrino and 
the charged lepton sectors. Indeed,
a crucial assumption is that the neutrino mass matrix $m_\nu$ and the combination
$m^\dagger_e m_e$, $m_e$ being the charged lepton mass matrix
\footnote{In our convention SU(2)$_L$ doublets are on the right of $m_e$.}
, are separately invariant under the 
subgroups $G_\nu$ and $G_e$ of $G_f$, respectively. 
We analyse possible mixing patterns independently from a specific model realisation,
and therefore we do neither specify the details of the symmetry breaking mechanism nor
 the transformation properties of fields under $G_f$ other than $l\sim 3$.
Neutrinos are assumed to be Majorana particles, which fixes $G_\nu$.
With a single generation, the only transformation of a Majorana neutrino leaving
invariant its mass term is a change of sign. If three generations are present, it can be shown \cite {lam} that
the appropriate invariance group of the neutrino sector is the product of two commuting parities,
the Klein group $Z_2\times Z_2$, allowing for an independent relative change of sign of any neutrino.
We assume $G_e$ to be abelian, since non-abelian subgroups would result in a complete or partial degeneracy of the
mass spectrum, a feature difficult to reconcile with the observed charged lepton mass hierarchy. We choose
$G_e=Z_3$ in our examples. This is actually the minimal choice of $G_e$ that can ensure three independent mass parameters
and allows to uniquely fix the mixing angles in the charged lepton sector, up to permutations.
 Since we are interested in minimal
realisations we require that the generators of the subgroups $G_\nu$ and $G_e$ give rise to
 the whole group $G_f$ and not only a subgroup of it,
which could otherwise be used as starting point instead of $G_f$. 

We call $\rho$ the three-dimensional representation of $G_f$ for the lepton doublets $l$ and the elements $g_{\nu i}$ of $G_\nu$ and $g_{e i}$ of $G_e$ are given by matrices $\rho(g_{\nu i})$ and $\rho(g_{e i})$, respectively. The invariance requirements read
\be
\rho(g_{\nu i})^T m_\nu~ \rho(g_{\nu i})=m_\nu~~~~~~~\mathrm{and}~~~~~~~\rho(g_{e i})^\dagger m^\dagger_e m_e \rho(g_{e i})=m^\dagger_e m_e~~~.
\label{AB}
\ee
Since $\rho$ is a unitary representation and $G_\nu$ and $G_e$ are abelian, there exist two unitary transformations $\Omega_\nu$ and $\Omega_e$ that
diagonalise the matrices $\rho(g_{\nu i})$ and $\rho(g_{e i})$
\be
\rho(g_{\nu i})_{diag}=\Omega_\nu^\dagger~ \rho(g_{\nu i})~ \Omega_\nu~~~~~~~\mathrm{and}~~~~~~~\rho(g_{e i})_{diag}=\Omega_e^\dagger~ \rho(g_{e i})~ \Omega_e~~~.
\label{CD}
\ee
Requiring eq.(\ref{AB}) to be fulfilled has as consequence that 
 $\Omega_\nu$ and $\Omega_e$ are also the transformations that diagonalise
$m_\nu$ and $m^\dagger_e m_e$, respectively. It follows that the lepton mixing matrix
is 
\be
U_{PMNS}=\Omega_e^\dagger\Omega_\nu~~~,
\ee
up to some redefinitions. 
Indeed $\Omega_e$ and $\Omega_\nu$ are defined up to a multiplication from the right by a diagonal matrix $K_{e,\nu}$ of phases, 
\be
\Omega_e \;\;\; \rightarrow \;\;\; \Omega_e K_e \;\;\;\;\;\; \mathrm{and} \;\;\;\;\;\; \Omega_\nu \;\;\; \rightarrow \;\;\; \Omega_\nu K_\nu \; .
\ee
The phase freedom associated with $K_e$ can be used to remove three phases from the combination
$\Omega_e^\dagger\Omega_\nu$, while the phase freedom associated with $K_\nu$ can be employed
to get real and positive eigenvalues of $m_\nu$. After that we are left with three physical phases in 
$\Omega_e^\dagger\Omega_\nu$: the Dirac $CP$ phase
$\delta_{CP}$ and the two Majorana phases. The latter cannot be predicted in our approach since the
 eigenvalues of $m_\nu$ remain unconstrained by the requirement in eq.(\ref{AB}).
The Dirac phase is instead determined by $\Omega_e^\dagger\Omega_\nu$. 
Similarly to the neutrino masses also the charged lepton masses remain free parameters
and thus we cannot fix the ordering of both rows and columns of $U_{PMNS}$. We use this freedom by choosing the order 
that allows mixing angles as close as possible to the experimental best fit values. Note that also
the exact value of $\delta_{CP}$ depends on the actual ordering of rows and columns and thus we can determine its value
only up to $\pi$.

From eqs. (\ref{AB},\ref{CD}) we see that the mixing matrix $U_{PMNS}$ is not sensitive to the overall sign of the 
matrices representing the elements of $G_\nu$ and $G_e$. Moreover if we replace the matrices
representing the elements of $G_\nu$ and $G_e$ by their complex conjugates, the mixing matrix
$U_{PMNS}$ becomes complex conjugated as well. Therefore representations $\rho$ and $\rho'$ 
that differ by an overall sign in the elements $g_{\nu i}$ and $g_{e i}$ and/or that are related by a complex conjugation
are not discussed separately.

Finally, concerning the choice of the flavour group $G_f$,
we consider two examples in which $G_f$ is $\Delta(96)$ and $\Delta(384)$, respectively.

Summarising, in our approach the lepton mixing originates from the misalignment 
of the remnant subgroups in neutrino and charged lepton sectors.
With the knowledge of $G_\nu$ and $G_e$ mixing angles and the phase $\delta_{CP}$ are predicted, 
while lepton masses and Majorana phases remain unconstrained.
 
\mathversion{bold}
\section{Mixing patterns with non-vanishing $\theta_{13}$}
\mathversion{normal}

We present two examples in which the lepton mixing matrix has non-vanishing $\theta_{13}$ and is determined as
outlined above. These examples are based on the two 
 flavour groups $G_f=\Delta(96)$ and $G_f=\Delta(384)$, respectively.
They belong to the series $\Delta(6 n^2)$, $n$ being a natural number, and are subgroups of the (inhomogeneous) modular
group $\Gamma$ which is isomorphic to the projective special linear group $PSL(2,Z)$. We note that the group $S_4$ with which TB mixing can be predicted 
is isomorphic to $\Delta(24)$. 

In our first example $G_f=\Delta(96)$ the generators $S$ and $T$ fulfill
the relations \cite{D96}
\footnote{In the Appendix we discuss the relation between $S$ and $T$ and the generators $a$, $b$, $c$ and $d$ chosen
in \cite{D96} to define the groups $\Delta(6n^2)$.}
\be
S^2=(ST)^3=T^8=\mathbb{1}~~~,~~~~~~~~(ST^{-1}ST)^3=\mathbb{1}~~~~,
\label{g8}
\ee
 The group has ten conjugacy classes: $\{E\}$, $3C_2$, $12C_2$, $32C_3$, $3C_4$, $3C'_4$, $6C_4$, $12C_4$, $12C_8$ and $12C'_8$,
where $E$ is the identity, the first number stands for the number of elements in the class and the index denotes the order of the 
elements. The group has 96 elements which are of order 1, 2, 3, 4 and 8, respectively. There are ten irreducible 
representations: two singlets,
one doublet, six triplets and one sextet. The character table of $\Delta(96)$ can be found in \cite{D96}. Two of the irreducible triplets are not faithful representations and are not used in our analysis. The remaining four triplets $\rho_i$ ($i=1...4$) are related among each other either by an overall change of sign of both
$\rho_i(S)$ and $\rho_i(T)$ and/or through complex conjugation. For this reason we 
restrict our analysis to a particular three-dimensional representation with
\vskip 0.2cm
\be
\rho(S)=\frac{1}{2}
\left(
\begin{array}{ccc}
0&\sqrt{2}&\sqrt{2}\\
\sqrt{2}&-1&1\\
\sqrt{2}&1&-1
\end{array}
\right)~~~~~~~~~~
\rho(T)=
\left(
\begin{array}{ccc}
e^{\frac{6\pi i}{4}}&0&0\\
0&e^{\frac{7\pi i}{4}}&0\\
0&0&e^{\frac{3\pi i}{4}}
\end{array}
\right)~~~.
\ee
\vskip 0.5cm
By choosing $G_\nu=Z_2\times Z_2$ and $G_e=Z_3$, we find seven distinct subgroups $Z_2\times Z_2$ and sixteen $Z_3$ subgroups, giving rise to 112 different possibilities for the lepton mixing matrix. Requiring that we generate the original group $\Delta(96)$ with the generators of $G_\nu$ and $G_e$ leaves us with 48 different combinations. As can be shown, all these are related by group transformations and
 thus produce the same $U_{PMNS}$, up to permutations of rows and columns and phase redefinitions.

\begin{figure}[h!]
\begin{center}
 \mbox{\epsfig{figure=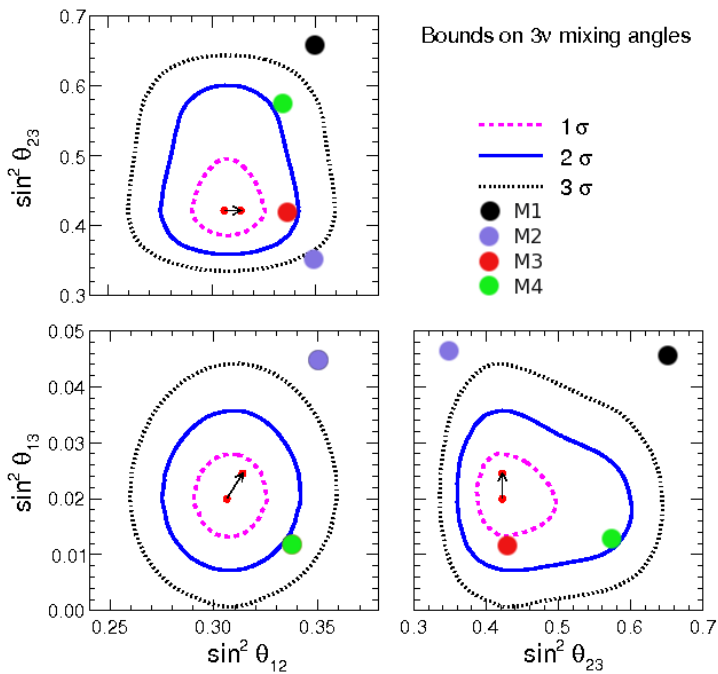,width=12.0cm}}
\end{center}
\caption{
Values of $\sin^2 \theta_{ij}$ for the four different mixing patterns \texttt{M1} (black), \texttt{M2} (violet), 
\texttt{M3} (red) and \texttt{M4} (green). The counters show the
$1 \sigma$ (pink dashed line), $2 \sigma$ (blue solid line) and $3 \sigma$ (black dotted line) levels and are taken from 
\cite{fogli}. The small dots indicate the best fit values of the mixing angles and the arrows the effect of the new estimates
of the reactor antineutrino flux. Note that in the $\sin^2 \theta_{12}$-$\sin^2 \theta_{13}$ plane the points of \texttt{M1} and \texttt{M2} as well
as of \texttt{M3} and \texttt{M4} lie on top of each other, since they only differ in the value of $\sin^2 \theta_{23}$.
}
\label{figure1}
\end{figure}

A possible choice for the generators of $G_\nu=Z_2\times Z_2$ and $G_e=Z_3$ is given by
\bea
G_\nu:&&\{ S,ST^4ST^4 \}\nn\\
G_e:&&ST~~~~~~~.
\eea
The absolute values $||U_{PMNS}||$ of the mixing matrix are 
\vskip 0.1cm
\be
\label{mixing1}
||U_{PMNS}||=
\frac{1}{\sqrt{3}}
\left(
\begin{array}{ccc}
\frac{1}{2}(\sqrt{3}+1) & 1 & \frac{1}{2}(\sqrt{3}-1) \\
\frac{1}{2}(\sqrt{3}-1) & 1 & \frac{1}{2}(\sqrt{3}+1) \\
   1                    & 1 & 1
\end{array}
\right)
\approx
\left(
\begin{array}{ccc}
 0.789 &  0.577 &0.211\\
 0.211 &  0.577 &0.789\\
 0.577  & 0.577 & 0.577
\end{array}
\right)~~~.
\ee
\vskip 0.3cm\noindent
With the ordering chosen in eq. (\ref{mixing1}), the mixing angles and the Dirac $CP$ phase read
\footnote{We use the conventions of \cite{pdg}.}
\bea
\sin^2\theta_{23}&=&\frac{5+2\sqrt{3}}{13}\approx 0.651\nn\\
\sin^2\theta_{12}&=&\frac{8-2\sqrt{3}}{13}\approx 0.349~~~~~~~~~~~~~~~~~~~({\tt M1})\nn\\
\sin^2\theta_{13}&=&\frac{2-\sqrt{3}}{6}\approx 0.045\nn\\
\delta_{CP}&=&\pi~~~.
\eea
If we exchange the second and third rows in $U_{PMNS}$ we have 
\bea
\sin^2\theta_{23}&=&\sin^2\theta_{12}=\frac{8-2\sqrt{3}}{13}\approx 0.349\nn\\
\sin^2\theta_{13}&=&\frac{2-\sqrt{3}}{6}\approx 0.045~~~~~~~~~~~~~~~~~~~~~~~~~~~~~~~~({\tt M2})\nn\\
\delta_{CP}&=&0~~~.
\eea
It is interesting to note that $CP$ is conserved in both cases.
The patterns {\tt M1} and {\tt M2} give rise to mixing angles which are compatible with the
present data, however only at roughly the $3\sigma$ level, as shown in figure \ref{figure1}. 
\vskip 0.5cm

In the second example $G_f=\Delta(384)$ $S$ and $T$ fulfill (see footnote 3)
\be
S^2=(ST)^3=T^{16}=\mathbb{1}~~~,~~~~~~~~(ST^{-1}ST)^3=\mathbb{1}~~~~.
\label{g16}
\ee
 The conjugacy classes are 24: 
$\{E\}$, $3C_2$, $24C_2$, $128C_3$, $3C_4$, $3C'_4$, $6C_4$, $24C_4$, $3C^i_8$, $6C^j_8$, $24C^k_8$, $24C^i_{16}$, ($i=1...4$), ($j=1...6$), ($k=1,2$).
The 384 elements of this group are of order 1, 2, 3, 4, 8 and 16, respectively.
There are 24 irreducible representations: two singlets,
one doublet, 14 triplets and seven sextets. Six triplets are unfaithful representations and are not considered here.
The remaining eight triplets can be divided into two sets containing four triplets each whose matrices $\rho(S)$ and $\rho(T)$
 for the generators $S$ and $T$ are related by an overall change of sign and/or through complex conjugation 
as in the previous example. As a consequence, we only need to consider the following two irreducible three-dimensional representations
\vskip 0.2cm
\be
\rho_1(S)=\frac{1}{2}
\left(
\begin{array}{ccc}
0&\sqrt{2}&\sqrt{2}\\
\sqrt{2}&-1&1\\
\sqrt{2}&1&-1
\end{array}
\right)~~~~~~~~~~
\rho_1(T)=
\left(
\begin{array}{ccc}
e^{\frac{14\pi i}{8}}&0&0\\
0&e^{\frac{5\pi i}{8}}&0\\
0&0&e^{\frac{13\pi i}{8}}
\end{array}
\right)~~~,
\label{r1}
\ee
and
\be
\rho_2(S)=\frac{1}{2}
\left(
\begin{array}{ccc}
0&\sqrt{2}&\sqrt{2}\\
\sqrt{2}&-1&1\\
\sqrt{2}&1&-1
\end{array}
\right)~~~~~~~~~~
\rho_2(T)=
\left(
\begin{array}{ccc}
e^{\frac{6\pi i}{8}}&0&0\\
0&e^{\frac{9\pi i}{8}}&0\\
0&0&e^{\frac{\pi i}{8}}
\end{array}
\right)~~~.
\label{r2}
\ee
\vskip 0.5cm
By choosing $G_\nu=Z_2\times Z_2$ and $G_e=Z_3$, we find 13 distinct subgroups $Z_2\times Z_2$ and 64 $Z_3$ subgroups, resulting in 832 different possibilities. Again, considering only those cases in which the generators of $G_\nu$ and $G_e$ give rise to the original group $\Delta(384)$, we are left with 384 combinations. It is easy to check that all these combinations are related by group transformations
and thus necessarily the same mixing matrix $U_{PMNS}$ is obtained, up to phase redefinitions and permutations of rows and columns. These statements hold for both representations $\rho_1$ and $\rho_2$ and moreover both of them give rise to the same mixing pattern.

 A possible choice for the generators of $G_\nu=Z_2\times Z_2$ and $G_e=Z_3$ is given by 
\bea
G_\nu:&&\{ S,ST^8ST^8 \}\nn\\
G_e:&&ST~~~~~~~.
\eea
The absolute values $||U_{PMNS}||$ of the mixing matrix are
\vskip 0.1cm
\bea
\label{mixing2}
||U_{PMNS}||
&=&\frac{1}{\sqrt{3}}
\left(
\begin{array}{ccc}
\frac{1}{2}\sqrt{4+\sqrt{2}+\sqrt{6}}&1& \frac{1}{2}\sqrt{4-\sqrt{2}-\sqrt{6}}\\
\frac{1}{2}\sqrt{4+\sqrt{2}-\sqrt{6}}&1&\frac{1}{2}\sqrt{4-\sqrt{2}+\sqrt{6}}\\
\sqrt{1-\frac{1}{\sqrt{2}}}&1&\sqrt{1+\frac{1}{\sqrt{2}}}
\end{array}
\right)\nn\\
&\approx&
\left(
\begin{array}{ccc}
0.810 & 0.577 & 0.107 \\
0.497 & 0.577 & 0.648\\
0.312 & 0.577 & 0.754
\end{array}
\right)~~~.
\eea
\vskip 0.3cm\noindent
With the ordering chosen in eq. (\ref{mixing2}), the mixing angles and the Dirac $CP$ phase are
\bea
\sin^2\theta_{23}&=&\frac{4-\sqrt{2}+\sqrt{6}}{8+\sqrt{2}+\sqrt{6}}\approx 0.424\nn\\
\sin^2\theta_{12}&=&\frac{4}{8+\sqrt{2}+\sqrt{6}}\approx 0.337~~~~~~~~~~~~~~~~~({\tt M3})\nn\\
\sin^2\theta_{13}&=&\frac{4-\sqrt{2}-\sqrt{6}}{12}\approx 0.011\nn\\
\delta_{CP}&=&0~~~.
\eea
If we exchange the second and third rows in $U_{PMNS}$ we have 
\bea
\sin^2\theta_{23}&=&\frac{4+2\sqrt{2}}{8+\sqrt{2}+\sqrt{6}}\approx 0.576\nn\\
\sin^2\theta_{12}&=&\frac{4}{8+\sqrt{2}+\sqrt{6}}\approx 0.337~~~~~~~~~~~~~~~~~({\tt M4})\nn\\
\sin^2\theta_{13}&=&\frac{4-\sqrt{2}-\sqrt{6}}{12}\approx 0.011\nn\\
\delta_{CP}&=&\pi~~~.
\eea
We observe that $CP$ is conserved in both cases.
The mixing pattern {\tt M3} is compatible with the present data at the $2 \sigma$ level, as can be seen
from figure \ref{figure1}, and provides
an excellent first order approximation in a theoretical description of the observed lepton mixing angles.

Notice that by taking $G_f=\Delta(24)\simeq S_4$ and by choosing $G_\nu=Z_2\times Z_2$ and $G_e=Z_3$ such that $G_f$
is generated by the elements of $G_\nu$ and $G_e$, the unique mixing pattern achieved with our approach is
 TB mixing, see also \cite{lam}.

It is interesting to note that both mixing matrices $U_{PMNS}$ whose absolute values are displayed in eqs. (\ref{mixing1}) 
and (\ref{mixing2}) can be brought into a form in which the second column has three entries equal to $1/\sqrt{3}$.
In doing so it becomes obvious that the results presented are related in a particular way to the TB mixing matrix
whose entries of the second column are usually defined to be all equal to $1/\sqrt{3}$ as well. Indeed, the TB mixing matrix
\be
U_{TB}=
\left(
\begin{array}{ccc}
\sqrt{\frac{2}{3}}&\frac{1}{\sqrt{3}}&0\\
-\frac{1}{\sqrt{6}}&\frac{1}{\sqrt{3}}&\frac{1}{\sqrt{2}}\\
-\frac{1}{\sqrt{6}}&\frac{1}{\sqrt{3}}&-\frac{1}{\sqrt{2}}
\end{array}
\right)
\ee
can be modified by a rotation in the 13 plane acting from the right
\be
U_{PMNS}=U_{TB} U_{13}(\alpha)~~~~~~~~~\mathrm{with}~~~~~~~~~~~
U_{13}(\alpha)=
\left(
\begin{array}{ccc}
\cos\alpha&0&\sin\alpha\\
0&1&0\\
-\sin\alpha&0&\cos\alpha
\end{array}
\right)~~~~.
\ee
\indent
It is immediate to show that, by taking $\alpha=-\pi/12$ and $\alpha=\pi/24$, the resulting mixing matrices are identical, in absolute value, to the matrices in eqs. (\ref{mixing1}) and (\ref{mixing2}), respectively.
Taking the opposite signs, $\alpha=\pi/12$ and $\alpha=-\pi/24$, we get the matrices with absolute values of the same form as in eqs. (\ref{mixing1}) and (\ref{mixing2}), respectively, with second and third rows exchanged. 
Such perturbations from TB mixing with $\alpha$ arbitrary have been already discussed in the literature \cite{TBmod,zee,lin}.
For generic $\alpha$, the mixing angles read 
\be
\sin^2\theta_{12}=\frac{1}{2+\cos 2\alpha}~,~~~\sin^2\theta_{23}=\frac{1}{2}-\frac{\sqrt{3}\sin 2\alpha}{4+2\cos 2\alpha}~,~~~~\sin^2\theta_{13}=\frac{2}{3}\sin^2\alpha~~.
\ee
For small $\alpha$, we can expand the results
\be
\sin^2\theta_{12}\approx\frac{1}{3} + \frac{2\alpha^2}{9}~,~~~\sin^2\theta_{23}\approx\frac{1}{2}-\frac{\alpha}{\sqrt{3}}~,~~~~\sin^2\theta_{13}\approx\frac{2 \alpha^2}{3}
\ee
showing that the deviation from the value of TB mixing of $\sin^2\theta_{12}$, the best measured quantity among the three mixing angles, is quadratic in $\alpha$, whereas the 
leading correction to $\sin^2 \theta_{23}=1/2$ is linear in $\alpha$.

\section{Conclusions}

Recent results of the T2K experiment and of a global fit of the neutrino oscillation data point
to non-vanishing $\theta_{13}$ at the 3$\sigma$ level. The best fit value of $\theta_{13}$ is around $0.15\div0.16$, smaller than the ones of the other angles, but much larger than $0.02$, the 1$\sigma$ experimental error on the solar angle $\theta_{12}$. 
If future data confirm this result, many models giving rise at LO to mixing patterns with vanishing $\theta_{13}$, such as TB mixing, 
become disfavoured, because corrections, expected in these models, generically lead to too small $\theta_{13}$.  
A particular elegant mechanism to produce simple mixing patterns is based on discrete flavour symmetries. The latter are broken in a 
non-trivial way and as a consequence give rise to mixing angles whose values only depend on the properties of the flavour symmetry, but
not on lepton masses. After the T2K data the natural
question is whether such symmetries still remain a valuable tool to describe
flavour mixing. One obvious possibility is to modify the existing models which lead to $\theta_{13}=0$ at LO,
by means of suitable perturbations to match the experimental data. 

In this note we have shown that
it is possible to predict a small, non-vanishing $\theta_{13}$ even in the absence of such perturbations,
in the framework of non-trivially broken discrete symmetries. The theory is invariant under a discrete flavour group $G_f$,
broken in such a way that the relevant mass matrices $m_\nu$ and $m_e^\dagger m_e$ have a residual invariance
under the subgroups $G_\nu$ and $G_e$, respectively. The lepton mixing matrix originates from the mismatch of these
two subgroups and from their specific embedding into $G_f$. By choosing $G_f=\Delta(6 n^2)$ ($n=4,8$), $G_\nu=Z_2\times Z_2$,
and $G_e=Z_3$ we find $\sin^2\theta_{13}=0.045(0.011)$ for $n=4(8)$. At the same time $\theta_{23}$ and $\theta_{12}$ are 
close to their experimental best fit values, especially in the case $n=8$ in which
  we find $\sin^2\theta_{23}\approx 0.424$ and $\sin^2\theta_{12}\approx 0.337$, see mixing pattern \texttt{M3}. The $CP$ violating phase $\delta_{CP}$ is predicted to be $0$ or $\pi$ so that $CP$ is conserved, at LO. Our proposed mixing patterns
are related to TB mixing in a simple way, namely they can be obtained
through a rotation by an angle $\alpha$, $\alpha=\pm\pi/12$ for $n=4$ and $\alpha=\pm\pi/24$ for $n=8$, respectively, in the 13 plane acting from the 
right on the TB mixing matrix.

Finally, we would like to mention that the presented results are part of a systematic investigation of finite subgroups of $PSL(2,Z)$, 
and that a comprehensive study is detailed in a future publication.

\section*{Acknowledgements}

We recognize that the work of FF has been partly supported by the European Programme "Unification in the LHC Era", contract PITN-GA-2009-237920 (UNILHC).
The work of RdAT  is part of the research program of the Dutch Foundation for Fundamental Research of Matter (FOM). RdAT acknowledges the hospitality of the University of Padova, where part of this research was completed. 

\section*{Appendix}

The groups $\Delta(6n^2)$ are non-abelian finite subgroups of $SU(3)$ 
of order $6n^2$. They are isomorphic to the semidirect product of  
$S_3$, the smallest non-abelian finite group, with
$Z_n\times Z_n$ \cite{D96,Bovier:1980gc},
\be
\Delta(6n^2)~\simeq~(Z_n\times Z_n)\rtimes S_3~~~.
\ee
They can be defined in terms of four generators $a$, $b$, $c$, $d$, satisfying
\be
a^3 ~=~ b^2 ~=~ (ab)^2 ~=~ c^n ~=~ d^n ~=~\mathbb{1}, 
\ee
\be
cd ~=~ dc ,
\ee
\be
\begin{array}{cccccc}
a c a^{-1}  &~=~&  c^{-1} d^{-1}, \quad
&a d a^{-1} &~=~&  c,             \\ 
b c b^{-1}  &~=~&  d^{-1},        
&b d b^{-1} &~=~&  c^{-1}.      \\    
\end{array}
\ee
The elements $a$ and $b$ are the generators of $S_3$ while
$c$ and $d$ generate $Z_n\times Z_n$. Here we show that the relations, found in eqs. (\ref{g8}) and (\ref{g16}) for $n=4$ and $n=8$
and given in terms of only two generators $S$ and $T$, indeed define the same group as those given for $a$, $b$, $c$ and $d$.
In the case of $n=4$, i.e. $\Delta(96)$, $S$ and $T$  are related to the generators above through
\be
\begin{array}{ll}
a=T^5ST^4~~~~~~~~~~~~~~&b=ST^2ST^5\\
c=ST^2ST^4~~~~~~~~~~~~~~&d=ST^2ST^6
\end{array}~~~~~~~.
\ee
For $\Delta(384)$ the relation is
\be
\begin{array}{ll}
a=T^{15}ST^8~~~~~~~~~~~~~~&b=ST^6ST^3\\
c=ST^2ST^4~~~~~~~~~~~~~~&d=ST^2ST^{14}
\end{array}~~~~~~~.
\ee
Note that if we apply the same similarity transformation to the elements $X_i$ given on the right-hand side of the
equations for $a$, $b$, $c$, $d$
\be
X_i\to g~X_i g^{-1}
\ee
with $g$ being an element of the group, we obtain an equally valid realisation of $a$, $b$, $c$, $d$.



\begin{thebibliography}{99}

\bibitem{abe11}
  K.~Abe {\it et al.}  [T2K Collaboration],
  arXiv:1106.2822 [hep-ex].

\bibitem{chooz}
M.~Apollonio {\it et al.}  [CHOOZ Collaboration],
  Eur.\ Phys.\ J.\  C {\bf 27} (2003) 331
  [arXiv:hep-ex/0301017].

\bibitem{minos}  
L. Whitehead [MINOS Collaboration], Recent results from MINOS, Joint Experimental-Theoretical Seminar (24 June
2011, Fermilab, USA). Websites: \verb1theory.fnal.gov/jetp1, \verb2http://www-numi.fnal.gov/pr_plots/2
  
\bibitem{theta13nonzero08}
 G.~L.~Fogli, E.~Lisi, A.~Marrone, A.~Palazzo, A.~M.~Rotunno,
  Phys.\ Rev.\ Lett.\  {\bf 101 } (2008)  141801.
  [arXiv:0806.2649 [hep-ph]].

\bibitem{fogli}
  G.~L.~Fogli, E.~Lisi, A.~Marrone, A.~Palazzo and A.~M.~Rotunno,
  arXiv:1106.6028 [hep-ph].

\bibitem{flux}
 T.~A.~Mueller {\it et al.},
  Phys.\ Rev.\  C {\bf 83} (2011) 054615
  [arXiv:1101.2663 [hep-ex]];
G.~Mention, M.~Fechner, T.~Lasserre, T.~A.~Mueller, D.~Lhuillier, M.~Cribier and A.~Letourneau,
  Phys.\ Rev.\  D {\bf 83} (2011) 073006
  [arXiv:1101.2755 [hep-ex]].

\bibitem{HPS}
P.~F.~Harrison, D.~H.~Perkins and W.~G.~Scott,
  Phys.\ Lett.\ B {\bf 530} (2002) 167
  [arXiv:hep-ph/0202074];
P.~F.~Harrison and W.~G.~Scott,
  Phys.\ Lett.\ B {\bf 535} (2002) 163
  [arXiv:hep-ph/0203209];
Z.~z.~Xing,
  Phys.\ Lett.\ B {\bf 533} (2002) 85
  [arXiv:hep-ph/0204049];
P.~F.~Harrison and W.~G.~Scott,
  Phys.\ Lett.\ B {\bf 557} (2003) 76
  [arXiv:hep-ph/0302025];
  arXiv:hep-ph/0402006.
  
\bibitem{reviews}
  G.~Altarelli, F.~Feruglio,
  Rev.\ Mod.\ Phys.\  {\bf 82 } (2010)  2701-2729.
  [arXiv:1002.0211 [hep-ph]];
  H.~Ishimori, T.~Kobayashi, H.~Ohki, Y.~Shimizu, H.~Okada, M.~Tanimoto,
  Prog.\ Theor.\ Phys.\ Suppl.\  {\bf 183 } (2010)  1-163.
  [arXiv:1003.3552 [hep-th]].

 \bibitem{gr}
   Y.~Kajiyama, M.~Raidal and A.~Strumia,
  Phys.\ Rev.\  D {\bf 76} (2007) 117301
  [arXiv:0705.4559 [hep-ph]];
  W.~Rodejohann,
  Phys.\ Lett.\  B {\bf 671} (2009) 267
  [arXiv:0810.5239 [hep-ph]];
  A.~Adulpravitchai, A.~Blum and W.~Rodejohann,
  New J.\ Phys.\  {\bf 11} (2009) 063026
  [arXiv:0903.0531 [hep-ph]];
L.~L.~Everett and A.~J.~Stuart,
  Phys.\ Rev.\  D {\bf 79}, 085005 (2009)
  [arXiv:0812.1057 [hep-ph]];
  F.~Feruglio and A.~Paris,
  JHEP {\bf 1103}, 101 (2011)
  [arXiv:1101.0393 [hep-ph]].
  
  \bibitem{hexa}
  C.~H.~Albright, A.~Dueck and W.~Rodejohann,
  Eur.\ Phys.\ J.\  C {\bf 70} (2010) 1099
  [arXiv:1004.2798 [hep-ph]].
  
\bibitem{TBmod}
X.~G.~He and A.~Zee,
  Phys.\ Lett.\  B {\bf 645} (2007) 427
  [arXiv:hep-ph/0607163];
W.~Grimus and L.~Lavoura,
  JHEP {\bf 0809}, 106 (2008)
  [arXiv:0809.0226 [hep-ph]];
 W.~Grimus, L.~Lavoura and A.~Singraber,
  Phys.\ Lett.\  B {\bf 686}, 141 (2010)
  [arXiv:0911.5120 [hep-ph]].
 
  \bibitem{zee}
  C.~H.~Albright, W.~Rodejohann,
  Eur.\ Phys.\ J.\  {\bf C62 } (2009)  599-608.
  [arXiv:0812.0436 [hep-ph]]; 
X.~G.~He and A.~Zee,
  arXiv:1106.4359 [hep-ph].
  
  \bibitem{ma}
  E.~Ma and D.~Wegman,
  arXiv:1106.4269 [hep-ph].

\bibitem{mpp_2011}
 S.~Morisi, K.~M.~Patel, E.~Peinado,
  arXiv:1107.0696 [hep-ph].

 \bibitem{lin}
  Y.~Lin,
  Nucl.\ Phys.\  B {\bf 824} (2010) 95
  [arXiv:0905.3534 [hep-ph]].
  
 \bibitem{meloni}
    G.~Altarelli, F.~Feruglio and L.~Merlo,
  JHEP {\bf 0905}, 020 (2009)
  [arXiv:0903.1940 [hep-ph]];
  D.~Meloni,
  arXiv:1107.0221 [hep-ph].
  
  \bibitem{lam}
  C.~S.~Lam,
  Phys.\ Lett.\  B {\bf 656} (2007) 193
  [arXiv:0708.3665 [hep-ph]];
  Phys.\ Rev.\ Lett.\  {\bf 101} (2008) 121602
  [arXiv:0804.2622 [hep-ph]];
  Phys.\ Rev.\  D {\bf 78} (2008) 073015
  [arXiv:0809.1185 [hep-ph]];
  Phys.\ Rev.\  D {\bf 83} (2011) 113002
  [arXiv:1104.0055 [hep-ph]].

\bibitem{D96}
  W. M. Fairbairn, T. Fulton, and W. H. Klink, J.\ Math.\ Phys.\ {\bf 5} (1964) 1038; 
  J.~A.~Escobar and C.~Luhn,
  J.\ Math.\ Phys.\  {\bf 50} (2009) 013524
  [arXiv:0809.0639 [hep-th]].

\bibitem{pdg}
K. Nakamura {\it et al.} [Particle Data Group], J.\ Phys.\ G {\bf 37} (2010) 075021 
and 2011 partial update for the 2012 edition in \verb3http://pdg.lbl.gov/3.
  
\bibitem{Bovier:1980gc}
  A.~Bovier, M.~Luling, D.~Wyler,
  J.\ Math.\ Phys.\  {\bf 22 } (1981)  1543.
   
\end{thebibliography}
\end{document}